\documentclass{kapproc}
\usepackage{procps}
\usepackage{graphicx}
\kluwerbib

\begin{document}

\articletitle
{Magnetic fields in the Milky Way and other spiral galaxies}

\chaptitlerunninghead{Magnetic fields in the Milky Way}

\author{Rainer Beck,\altaffilmark{}}

\affil{\altaffilmark{}MPI f\"ur Radioastronomie, Auf dem H\"{u}gel 69,
53121 Bonn, Germany}

\begin{abstract}

The average strength of the {\it total} magnetic field in the Milky Way,
derived from radio synchrotron data under the energy equipartition assumption,
is $6\mu$G locally and $\simeq10\mu$G at 3~kpc Galactic radius.
Optical and synchrotron polarization data yield a strength of the local
{\it regular} field of $\simeq4\mu$G (an upper limit if anisotropic fields
are present), while pulsar rotation measures give $\simeq1.5\mu$G (a
lower limit if small-scale fluctuations in regular field strength and in
thermal electron density are anticorrelated). In spiral arms of external
galaxies, the total [regular] field strength is up to $\simeq35\mu$G
[$\simeq15\mu$G]. In nuclear starburst regions the total field reaches
$\simeq50\mu$G. -- -- 
Little is known about the global field structure in the Milky Way.
The local regular field may be part of a ``magnetic arm'' between the
optical arms, a feature that is known from other spiral galaxies.
Unlike external galaxies, rotation measure data indicate several global
field reversals in the Milky Way, but some of these could be due to
field distortions. The Galaxy is surrounded by a thick radio disk of
similar extent as around many edge-on spiral galaxies. While the 
regular field of the local disk is of even symmetry with respect to the plane 
(quadrupole), the regular field in the inner Galaxy or in the halo may be of 
dipole type. The Galactic center region hosts highly regular fields 
of up to milligauss strength which are oriented perpendicular to the plane.

\end{abstract}


\section{Motivation}

Magnetic fields are a major agent in the interstellar medium. They 
contribute significantly to the total pressure which balances the ISM
against gravitation. They affect the gas flows in spiral arms and
around bars. Magnetic fields are essential for the onset of star 
formation as they enable the removal of angular momentum from the 
protostellar cloud during its collapse. MHD turbulence distributes 
energy from supernova explosions within the ISM. Magnetic reconnection 
is a possible heating source for the ISM and halo gas. Magnetic fields 
also control the density and distribution of cosmic rays in the ISM.

\section{Observing magnetic fields}

{\it Polarized emission} at optical, infrared, submillimeter and radio
wavelengths is the clue to interstellar magnetic fields. Radio continuum 
emission at centimeter wavelengths, emitted by cosmic-ray electrons in the 
interstellar medium, by pulsars and by background quasars,
has higher degrees of polarization than in the other spectral ranges
and provides the most extensive and reliable information on large-scale
interstellar magnetic fields in our Galaxy (Heiles 1996, Han et al. 1999a)
and about 70 external galaxies (see list in Beck 2000).
The observable degree of polarization
is reduced by the contribution of unpolarized thermal emission
which may dominate in star-forming regions, by Faraday depolarization
(Sokoloff et al. 1998) and by geometrical depolarization within the beam.
The orientation of polarization vectors is changed in a magneto-ionic
medium by {\it Faraday rotation} which is generally small
below about $\lambda6$~cm so that the $\bf{B}$--vectors (i.e. the
observed $\bf{E}$--vectors rotated by $90^{\circ}$) directly trace
the {\it orientation\/} of the regular (or anisotropic) fields in the
sky plane. 

Polarization angles are ambiguous by $\pm 180^{\circ}$ and hence
insensitive to field reversals. Compression or stretching of turbulent 
fields generates incoherent anisotropic fields which reverse direction 
frequently within the telescope beam, so that Faraday rotation is small 
while the degree of polarization can be high. Strong Faraday rotation 
is a signature of coherent regular fields, and the sense of rotation 
reveals the {\it sign} of the field.

\section{Magnetic field strengths}

The average strength of the total $\langle B_{\mathrm{t},\perp}\rangle$
and the resolved regular field $\langle B_{\mathrm{reg},\perp}\rangle$
in the plane of the sky can be derived from the total and polarized
radio synchrotron intensity, respectively, if energy-density
equipartition between cosmic rays and magnetic fields is assumed
(see Beck et al. 1996 for details).

In our Galaxy the accuracy of the equipartition assumption can be
tested, because we have independent information 
about the local cosmic-ray energy density from in-situ measurements
and about their radial distribution from $\gamma$-ray data.
Combination with the radio synchrotron data yields a local
strength of the total field $\langle B_\mathrm{t}\rangle$ of $6\mu$G
(Strong et al. 2000), the same value as derived from
energy equipartition (Berkhuijsen, in Beck 2001).
Even the radial scale length of the equipartition
field of $\simeq12$~kpc (Fig.~1) is similar to that in Strong et al. (2000).
Near the Galactic center the field strength reaches 1~mG
(Reich 1994, Yusef-Zadeh et al. 1996).

\begin{figure}
\centerline{\includegraphics[width=8.5cm]{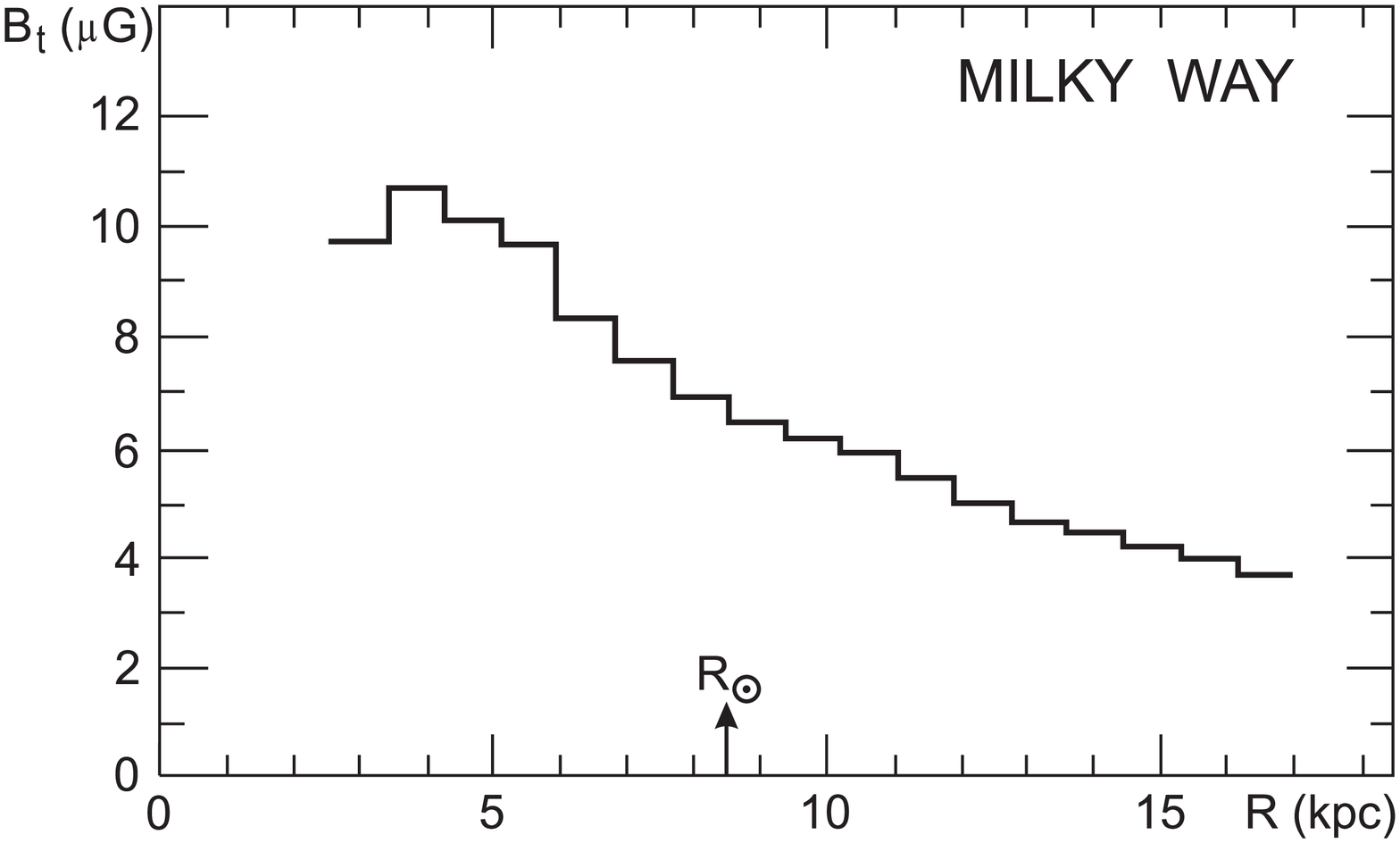}}
\vspace{0.2cm}
\caption{ Strength of the total magnetic field in the Galaxy,
averaged from the deconvolved surface brightness of the
synchrotron emission at 408~MHz (Beuermann et al. 1985),
assuming energy equipartition between magnetic field and cosmic
ray energy densities. The accuracy is about 30\%. 
The Sun is assumed to be located at R=8.5~kpc. 
(From Berkhuijsen, in Beck 2001)}
\end{figure}

The mean equipartition strength of the total field for a sample of 74 
spiral galaxies (Niklas 1995) is $\langle B_\mathrm{t}\rangle\simeq9\mu$G.
Radio-faint galaxies like M\,31 and M\,33 have
$\langle B_\mathrm{t}\rangle\simeq6\mu$G, while $\simeq15\mu$G 
is typical for grand-design galaxies like M\,51, M\,83 and NGC~6946. 
In the prominent spiral arms of M\,51 the total field strength is
30--35$\mu$G (Fletcher et al., this volume). Nuclear starburst
regions host fields up to $\simeq50\mu$G strength 
(Klein et al. 1988, Beck et al. 1999).

Synchrotron polarization observations in the local Galaxy 
imply a ratio of regular to total field strengths of
$<B_{\mathrm{reg}}/B_\mathrm{t}>\simeq0.6$
(Berkhuijsen 1971, Brouw \& Spoelstra 1976, Heiles 1996).
The total radio emission along the local spiral arm requires that
$<B_\mathrm{reg}/B_\mathrm{t}>$=0.6--0.7
(Phillipps et al. 1981). For $\langle
B_\mathrm{t}\rangle=6\pm2\,\mu$G these results give $4\pm1\,\mu$G
for the local regular field component. Note that equipartition values
for $<B_\mathrm{reg}>$ may overestimate the strength of the coherent
regular field if reversals or anisotropic turbulent fields are present.

Rotation measure and dispersion measure data of pulsars give an 
average strength of the local regular field of 
$<B_{\mathrm{reg}}>=1.4\pm0.2\mu$G (Rand \& Lyne 1994, Han \& Qiao 1994, 
Indrani \& Deshpande 1998), less than the equipartition estimate.
$<B_\mathrm{reg}>$ derived from pulsar data is underestimated if 
small-scale fluctuations in field strength
and in electron density are anticorrelated, as expected for local
pressure equilibrium (Beck et al. 2003).
In the inner Norma arm, the average strength of the regular field is 
$4.4\pm0.9\mu$G (Han et al. 2002).

The strength of regular fields $B_\mathrm{reg}$ in
spiral galaxies (observed with a spatial resolution of a few
100~pc) is typically 1--5$\mu$G.  Exceptionally strong regular fields
are detected in the interarm region of NGC~6946 of up to $\simeq 13\mu$G 
(Beck \& Hoernes 1996, see Fig.~3)) and 
$\simeq15\mu$G at the inner edge of the inner spiral arms
in M\,51 (Fletcher, this volume).

In spiral arms of external galaxies the regular field is generally
weaker and the tangled (unresolved) field is stronger
due to turbulent gas motion in star-forming regions and the expansion of
supernova remnants. In interarm regions the regular field is generally
stronger than the tangled field.

\section{Energy densities in the ISM}

The relative importance of various competing forces in the interstellar
medium can be estimated by comparing the corresponding energy densities.
In the local Milky Way, the energy densities of 
turbulent gas motions, cosmic rays, and magnetic fields are similar 
(Boulares \& Cox 1990). Global studies are possible in external galaxies
like NGC~6946 (Fig.~2). The energy density of the total equipartition
magnetic field ($B^2_\mathrm{t}/8\pi$) is derived from the map of synchrotron
emission of Walsh et al. (2002), the thermal energy density
(${3\over2} \langle n_\mathrm{e} \rangle k T$) of the warm ionized gas 
($T\simeq 10^4$~K) from the map of thermal radio emission $I_{th}$ of 
Walsh et al., using a constant volume filling factor of 5\% (see however 
Mitra et al., this volume), and the thermal energy density 
of the total neutral gas (molecular + atomic) from the CO map of Walsh et al.
and the HI map of Kamphuis \& Sancisi (1993), assuming for simplicity a 
constant scale height of the disk of 100~pc and $T\simeq 50$~K.
For the kinetic energy density (${1\over2} \rho v^2$) of the turbulent
motion of the total neutral gas the turbulent velocity of the neutral
gas is assumed to be $v_\mathrm{turb}=7$~km/s, as for the
neutral gas in our Galaxy (Boulares \& Cox 1990, Kalberla \& Kerp 1998).

According to Fig.~2 the energy density of the ionized gas $E_{th}$
in NGC~6946 is one order of magnitude smaller than that of the 
magnetic field $E_{magn}$ (similar to the results for the Milky Way
by Boulares \& Cox 1990),
i.e. the ISM is a low-$\beta$ plasma. However, $\beta=E_{th}/E_{magn}$ 
may be underestimated if there is a contribution of hot gas or if the filling 
factor $f_V$ of the diffuse ionized gas is larger than the assumed 5\% 
(because $\langle n_\mathrm{e} \rangle \propto (I_{th} f_V)^{0.5}$).

\begin{figure}
\centerline{\includegraphics[width=8.5cm]{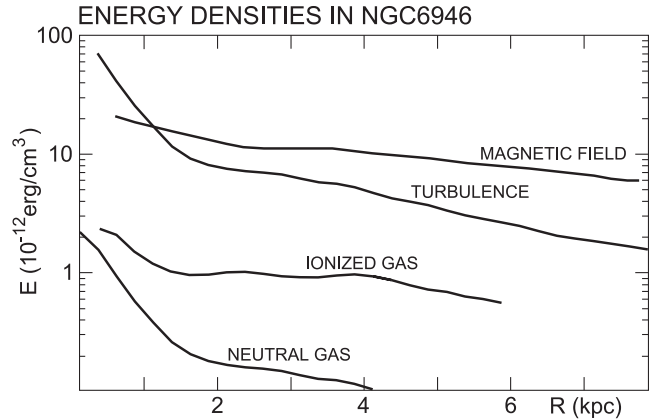}}
\vspace{0.2cm}
\caption{ Energy densities and their radial variations in the spiral
galaxy NGC~6946.}
\end{figure}

In the inner parts of NGC~6946 the energy densities of the total magnetic 
field and turbulent gas motion are similar (Fig.~2), but the field 
dominates in the outer parts due to the large radial scale length of the 
total field energy ($\simeq$8~kpc), compared to the scale length of 
about 3~kpc for the neutral density. This seems to be in conflict with
turbulent generation of interstellar magnetic fields.
Radial diffusion of the magnetic field (Priklonsky et al. 2000), field 
connections through the wind-driven halo (Breitschwerdt et al. 2002),
or a supra-equipartition turbulent dynamo (Belyanin et al. 1993)
are possible explanations.

NGC~6946 rotates with $v_\mathrm{rot}\simeq170$~km/s
so that the rotational energy density of the neutral gas is
$\simeq500\times$ larger than that of the turbulent motion. However,
in the outermost parts of galaxies the magnetic field energy density
may reach the level of global rotational gas motion and affect
the rotation curve, as proposed by Battaner \& Florido (2000).
Field strengths in the outer parts of galaxies can be measured
by Faraday rotation of polarized background sources.
Han et al. (1998) found evidence for regular fields in M\,31
at 25~kpc radius of similar strength as in the inner disk.
More detailed studies in a number of galaxies are required.

Magnetic fields seem to play a major and possibly even dominant role
in ISM physics, affecting gas flows, cloud collisions and the formation of
spiral arms. Strongly tangled fields may provide a source for gas heating
by reconnection.

\begin{figure}[htb]
\center
\centerline{\includegraphics[bb = 39 191 567 639,width=9cm]{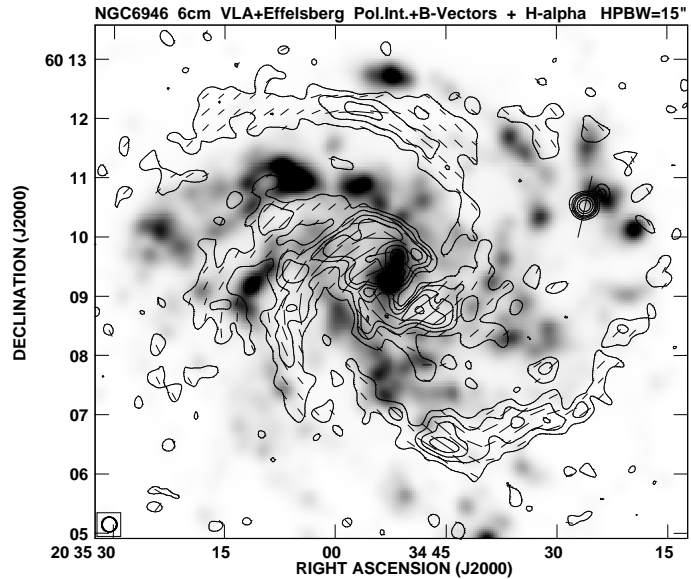}}
\vspace{0.2cm}
\caption{Polarized radio intensity (contours) and
$\mathbf{B}$--vectors of polarized
intensity of NGC~6946 at $15^{\prime\prime}$ resolution, combined
from VLA and Effelsberg observations at $\lambda$6~cm. The grey-scale
image shows the H$\alpha$ emission (kindly provided by A.~Ferguson),
smoothed to the same resolution. (From \protect{Beck \& Hoernes 1996})
}
\end{figure}

\section{Large-scale field structures}

The Sun is located between two spiral arms, the Sagittarius/Carina
and the Perseus arms. The mean pitch angle of the spiral arms is 
$\simeq -18^{\circ}$ for the stars and $-13^{\circ}\pm1^{\circ}$ for 
all gas components (Vall{\'e}e 1995, 2002). Starlight polarization 
and pulsar $RM$ data give a significantly smaller pitch
angle ($-8^{\circ}\pm1^{\circ}$) for the local magnetic field
(Heiles 1996, Han \& Qiao 1994, Indrani \& Deshpande 1998,
Han et al. 1999a). The local field may form a {\it magnetic arm} 
located between two optical arms, with a smaller pitch angle.

For external galaxies, maps of the total radio emission and ISOCAM maps
of the mid-infrared dust emission reveal a surprisingly close
connection (Frick et al. 2001b, Walsh et al. 2002). Strongest {\it total}
fields generally coincide with highest emission from dust and gas in 
the spiral arms. This suggests a coupling of the tangled magnetic field 
to the warm dust mixed with cool gas. The {\it regular} field runs 
parallel to the spiral arms, though its ridge line is generally offset and, 
in some galaxies, forms {\it magnetic spiral arms} between the gaseous arms
(Fig.~3) or across the arms, like in NGC~3627 (Soida et al. 2001).
In galaxies with strong density waves (like M\,51) the regular field 
fills the whole interarm space, but is strongest at the positions of 
the dust lanes on the inner edge of the spiral arms (Fletcher, this volume).
The low arm-interarm contrast is in conflict with classical 1-D shock
compression and needs more sophisticated 3-D modelling
(Martos \& Cox 1998, G\'omez \& Cox, this volume).

The observation of large-scale patterns in Faraday rotation measures
in the Milky Way and in external galaxies proves that a significant 
fraction of the field has a coherent direction and hence
is not generated by compression or stretching
in gas flows. The {\it dynamo} is able to generate and
preserve coherent magnetic fields of spiral shape (Beck et al. 1996,
Ferri{\`e}re, this volume).
Most observed field patterns require the superposition of several 
dynamo modes (Beck 2000).

A puzzling property of the Galactic magnetic field is
the existence of several {\it reversals} (Rand \& Lyne 1994,
Vall{\'e}e 1995, Han et al. 1999a, Frick et al. 2001a).
To account for several reversals along
Galactic radius, a {\it bisymmetric} magnetic spiral with a small pitch
angle ($\simeq -7^{\circ}$) has been proposed (Han \& Qiao1994,
Indrani \& Deshpande 1998, Han et al. 2002).

It is striking that only very few field reversals have been detected in
spiral galaxies. High-resolution maps of Faraday rotation, which measure
the $RMs$ of the diffuse polarized synchrotron emission, are
available for a couple of spiral galaxies. A dominating bisymmetric field 
structure was found only in M\,81 (Krause et al. 1989). 
The disk fields of M\,51 and NGC~4414 can be described by a mixture of
dynamo modes which would appear like a reversal to an observer located 
within the disk (Berkhuijsen et al. 1997, Soida et al. 2002).
In NGC~2997 a field reversal between the disk and the central region
occurs at about 2~kpc radius (Han et al. 1999b).
However, no multiple reversals along radius, like those in the
Milky Way, were found so far in any external galaxy.

The discrepancy between Galactic and extragalactic data may be due to
the different volumes traced by the observations. Results in the Galaxy
are based on pulsar $RMs$ which trace the warm ionized medium near the
plane, while extragalactic $RMs$ are averages over the whole thick disk.
Some of the Galactic
reversals may not be of galactic extent, but due to local field distortions 
or loops of the anisotropic turbulent field. Pulsar $RMs$ around a star 
formation complex indeed revealed a field distortion which may mimic the
reversal claimed to exist in the direction of the Perseus arm 
(Mitra et al. 2003). Many reversals on small scales are visible
in the $RM$ maps obtained from the diffuse Galactic synchrotron emission
(Haverkorn et al. 2003a, 2003b). $RM$ data at high frequencies are 
needed to obtain a clearer picture of the field structure.

\section{Vertical fields}

Edge-on galaxies possess thick radio disks (also called halos)
of 1--3~kpc scale height (Lisenfeld \& Dahlem, this volume), 
and the observed field orientations are mainly parallel
to the disk (Dumke et al. 1995). A prominent exception is
NGC~4631 with the brightest and largest halo observed so far,
composed of vertical magnetic spurs connected to
star-forming regions in the disk (Golla \& Hummel 1994).
NGC~5775 is an intermediate case with parallel and vertical
field components (T{\"u}llmann et al. 2000).
The magnetic energy density in the halo of, e.g. M\,83, exceeds 
that of the hot gas (Ehle et al. 1998), so that halo magnetic 
fields are important for the formation of a galactic wind.
Magnetic reconnection is a possible heating source
(Birk et al. 1998).

The vertical full equivalent thickness of the thick radio disk of
the Milky Way is $3.0 \pm 0.2$~kpc near the
Sun (Beuermann et al. 1985, scaled to a distance to the
Galactic center of 8.5~kpc) which corresponds to an exponential
scale height of $h_\mathrm{syn}\simeq1.5$~kpc. In case of equipartition
the scale height of the total field is $\simeq 4$ times larger than
that of the synchrotron disk, i.e.  $h_{B_\mathrm{t}} \simeq 6$~kpc.
The local Galactic field is oriented mainly parallel
to the plane, with a weak vertical component of $B_z\simeq0.2\mu$G
(Han \& Qiao 1994).

Dynamo models predict the preferred generation of quadrupole fields
where the toroidal component has the same sign above and below the plane.
$RMs$ of extragalactic sources and  pulsars reveal
no reversal near the plane for Galactic longitudes
$\,l=90^{\circ} - 270^{\circ}$. Thus the local field is part
of a large-scale symmetric (quadrupole) field structure.
Towards the inner Galaxy ($l=270^{\circ} - 90^{\circ}$) 
the signs are opposite above and below the
plane. This may indicate a global antisymmetric (dipole) mode
in the inner Galaxy or in the halo (Han et al. 1997) with a poloidal 
field component perpendicular to the plane, which may explain the
strong vertical fields observed near the Galactic center (see Sect.~2).

In external galaxies the vertical field symmetry could not be
determined yet. A possible dominance of inward-directed
radial field components may give evidence for 
preferred quadrupole-type fields (Krause \& Beck 1998).

\section{Small-scale field structures}

Major progress in detecting small structures has been achieved with 
decimeter-wave polarization observations in the Milky Way 
(Duncan et al. 1997, 1999, Uyan{\i}ker et al. 1998, 1999, 2003, 
Gaensler et al. 2001, Uyan{\i}ker \& Landecker 2002, Haverkorn et al. 2003a,b). 
A wealth of structures on pc and sub-pc scales has been discovered:
filaments, canals, lenses, and rings. Their common property is to appear 
only in the maps of polarized intensity, but not in total intensity. The 
interpretation is hampered by several difficulties. Firstly, large-scale 
emission in Stokes parameters Q and U is missing in interferometric and 
even in single-dish maps so that the polarized intensities and angles
can be distorted severely. Secondly, the wavelengths of these 
polarization surveys are rather long, so that strong depolarization 
of background emission in the foregound {\it Faraday screen}
may lead to apparent structures like {\it Faraday ghosts} (Shukurov
\& Berkhuijsen 2003). On the other hand, such features carry valuable 
information about the turbulent ISM in the Faraday screen.

Another effect of Faraday depolarization is that, especially at decimeter
wavelengths, only emission from nearby regions may be detected.
The ISM is not always transparent for polarized radio waves, and the 
opacity varies strongly with wavelength and position. (This is why 
{\it polarization horizon} seems a less appropriate expression.) The wavelength
dependence of Faraday depolarization allows {\it Faraday tomography} of
different layers if maps at different (nearby) wavelengths are combined.
More interesting results can be expected in this field.

\section{Discussion}

\noindent
{\it Boulanger:} The correlation between $B_{\mathrm{t}}$ and 
mid-IR emission measured by ISO: Isn't it a correlation between
$B_{\mathrm{t}}$ and the power input from star formation rather
than with cool gas?\\

\noindent
{\it Beck:} Yes, there must be some physical interaction between magnetic 
fields and star formation which we don't understand yet. The correlation
between radio continuum and H$\alpha$ emission is much
worse than between radio and mid-IR. A connection between magnetic
fields and gas+dust clouds heated by star formation seems
the most promising interpretation.\\

\noindent
{\it Rand:} Is there evidence for differing pitch angles of the
fields between arm and interarm regions in external spirals?\\

\noindent
{\it Beck:} Yes, the pitch angles of the field in NGC~6946 are 
smaller in the interarm regions ($\simeq15^{\circ}$) than in
the spiral arms ($\simeq30^{\circ}$).\\

\noindent
{\it Franco:} The north-south field reversal in the inner Galaxy
may imply that there is a neutral (current) sheet near the plane.\\

\noindent
{\it Beck:} The regular field becomes weak towards the plane due
to strong field tangling so that a neutral sheet will probably not
develop.\\

\noindent
{\it Mac Low:} If we see turbulent field, something must be tangling it.
Magnetic tension straightens field lines if nothing opposes it.
This appears to contradict the observation that field energy exceeds
turbulent energy.\\

\noindent
{\it Beck:} The {\it total} field dominates in the outer parts of 
NGC~6946, not the turbulent field. As the polarized emission decreases 
more slowly with radius than the unpolarized emission, the regular field 
becomes more important far out. Furthermore, the unpolarized synchrotron 
emission traces the unresolved field, consisting of field tangled 
on relatively large scales (though still unresolved by the telescope 
beam) and field which is really turbulent on small scales.
Hence, the energy of the turbulent field makes only some fraction of 
the total field energy and does not necessarily exceed the turbulent 
energy of the gas.\\

\noindent
{\it G{\'o}mez:} What if the $B$-field is too tangled, weaved on itself, 
knotted? You could have a dominanting fluctuating field
that does not need a constant driving.\\ 

\noindent
{\it Mac Low:} A knotted field is very unlikely.
\end{document}